\catcode`\@=11

\@addtoreset{equation}{section}

\def\@normalsize{\@setsize\normalsize{15pt}\xiipt\@xiipt
\abovedisplayskip 14pt plus3pt minus3pt%
\belowdisplayskip \abovedisplayskip
\abovedisplayshortskip  \z@ plus3pt%
\belowdisplayshortskip  7pt plus3.5pt minus0pt}

\def\small{\@setsize\small{13.6pt}\xipt\@xipt
\abovedisplayskip 13pt plus3pt minus3pt%
\belowdisplayskip \abovedisplayskip
\abovedisplayshortskip  \z@ plus3pt%
\belowdisplayshortskip  7pt plus3.5pt minus0pt

\def\@listi{\parsep 4.5pt plus 2pt minus 1pt
            \itemsep \parsep
            \topsep 9pt plus 3pt minus 3pt}}

\def\underline#1{\relax\ifmmode\@@underline#1\else
        $\@@underline{\hbox{#1}}$\relax\fi}
\@twosidetrue

\catcode`@=12

\catcode`\@=11

\catcode`\@=12

\relax

\def\figcap{\section*{Figure Captions\markboth
        {FIGURECAPTIONS}{FIGURECAPTIONS}}\list
        {Fig. \arabic{enumi}:\hfill}{\settowidth\labelwidth{Fig. 999:}
        \leftmargin\labelwidth
        \advance\leftmargin\labelsep\usecounter{enumi}}}
 \relax
\def\tablecap{\section*{Table Captions\markboth
        {TABLECAPTIONS}{TABLECAPTIONS}}\list
        {Table \arabic{enumi}:\hfill}{\settowidth\labelwidth{Table 999:}
        \leftmargin\labelwidth
        \advance\leftmargin\labelsep\usecounter{enumi}}}
 \relax
\def\reflist{\subsubsection*{References\markboth
        {REFLIST}{REFLIST}}\list
        {[\arabic{enumi}]\hfill}{\settowidth\labelwidth{[999]}
        \leftmargin\labelwidth
        \advance\leftmargin\labelsep\usecounter{enumi}}}
 \relax

\catcode`\@=11

\def\FERMIPUB{}

\def\ps@headings{\def\@oddfoot{}\def\@evenfoot{}
\def\@oddhead{\hbox{}\hfill
        \makebox[.5\textwidth]{\raggedright\ignorespaces --\thepage{}--
        \hfill {\rm FERMILAB--Pub--\FERMIPUB}}}
\def\@evenhead{\@oddhead}
\def\subsectionmark##1{\markboth{##1}{}}
}

\ps@headings

\relax

\newskip\humongous \humongous=0pt plus 1000pt minus 1000pt

\newif\ifdtup

\def\beq{\begin{equation}}
\def\eeq{\end{equation}}

\def\beqn{\begin{eqnarray}}
\def\eeqn{\end{eqnarray}}
\relax

\def\dotx{\dotx{\dot\overline{x}}}

\documentstyle[art12]{article}
\def\today{\number\day\space
     \ifcase\month\or
       January\or February\or March\or April\or May\or June\or
       July\or August\or September\or October\or November\or December\fi
     \space\number\year}
\begin{document}
\begin{titlepage}
\begin{flushright}
Z\"urich University ZU-TH 20/96\\
\end{flushright}
\vfill
\vskip 1cm
\begin{center}
{\large\bf MICROLENSING IMPLICATIONS FOR HALO DARK MATTER
\footnote{
To appear in the proceedings of the Journ\'ees Relativistes 96
(Ascona, 25-30 May 1996), which will be published in Helv. Phys. Acta}}\\
\vskip 0.5cm
{\bf Philippe~Jetzer}\\
\vskip 0.5cm
Paul Scherrer Institute, Laboratory for Astrophysics,
CH-5232 Villigen PSI, and\\
Institute of Theoretical Physics, University of Z\"urich,
Winterthurerstrasse 190,\\
CH-8057 Z\"urich, Switzerland\\
\end{center}
\vskip 0.5 cm
\begin{center}
Abstract
\end{center}
\begin{quote}
The French collaboration EROS and the American-Australian
collaboration MACHO have reported the observation of 
altogether $\sim$ 10 microlensing
events by monitoring during several years the brightness of millions
of stars in the Large Magellanic Cloud. 
In particular the MACHO team announced the discovery of 
8 microlensing candidates by analysing their first 2 years
of observations. This would imply that the halo dark matter
fraction in form of MACHOs 
(Massive Astrophysical Compact Halo Objects)
is of the order of 45-50\%. 
The most accurate way
to get information on the mass of the MACHOs
is to use the method of mass moments.
For the microlensing events detected so far
by the MACHO collaboration in the Large Magellanic Cloud
the average mass turns out to be 0.27$M_{\odot}$.
\end{quote}
\vfill
\end{titlepage}
\newpage
\section{Introduction}

It has been pointed out by Paczy\'nski \cite{kn:Paczynski} that microlensing 
allows the detection of MACHOs located in the galactic halo in the mass
range \cite{kn:Derujula1}  
$10^{-7} < M/M_{\odot} <  1$.
In September 1993 the French collaboration EROS \cite{kn:Aubourg}
announced the discovery of 2 microlensing candidates
and the American--Australian
collaboration MACHO of one candidate \cite{kn:Alcock}.
In the meantime the MACHO team reported the observation of
altogether 8 events
(one of which is a binary lensing event) analyzing
2 years of their data
by monitoring about 8.5 million 
of stars in the Large Magellanic Cloud (LMC) \cite{kn:Pratt}. 
Their analysis leads to an optical depth of $\tau=2.9^{+1.4}_{-0.9}
\times 10^{-7}$ and correspondingly to a halo MACHO fraction of the
order of 45-50\% and an average mass $0.5^{+0.3}_{-0.2} M_{\odot}$, 
under the assumption of a standard spherical halo model.
It may well be that there is also a contribution of events due
to MACHOs located in the LMC itself or in a thich disk of our galaxy,
the corresponding optical depth is estimated to be
$\tau=5.4 \times 10^{-8}$
\cite{kn:Pratt}.
Moreover, the
Polish-American team OGLE \cite{kn:Udalski}, the MACHO  
\cite{kn:MACHO} and the French DUO \cite{kn:Alard} collaborations 
found altogether more than $\sim$ 100  
microlensing events by monitoring stars located in the galactic bulge.
The inferred optical depth for the bulge turns out to be
higher than previously thought.
 
An important issue is 
the determination of the mass of the MACHOs that
acted as gra\-vi\-tational lenses as well as the fraction of halo dark
matter in form of MACHOs.
The most appropriate way to compute the average mass and other
important information is to use
the method of mass moments developed by De R\'ujula et al. \cite{kn:Derujula},
which will be briefly presented in section 3.\\ 

\section{Most probable mass for a single event}

First, we compute the probability $P$ that a microlensing
event of duration T
and maximum amplification $A_{max}$ be produced by a MACHO
of mass $\mu$ (in units of $M_{\odot}$).
Let $d$ be the distance of the MACHO from the line of
sight between the observer and a star in the LMC, t=0 the instant
of closest approach and $v_T$ the MACHO velocity in the transverse
plane. The magnification $A$ as a function of time is calculated using
simple geometry and is given by
\begin{equation}
A(t)=\frac{u^2+2}{u(u^2+4)^{1/2}}~, ~~~{\rm where}~~~
u^2=\frac{d^2+v_T^2 t^2}{R_E^2}~. 
\label{eqno:1}
\end{equation}
$R_E$ is the Einstein radius
which is
$R_E^2=\frac{4GMD}{c^2}x(1-x)=r_E^2 \mu x(1-x)$
with $M = \mu M_{\odot}$ the MACHO mass 
and $D~ (xD)$ the distance
from the observer to the source (to the MACHO). $D = 55$ kpc is the distance
to the LMC and $r_E=3.17 \times 10^9$~ km.
We use here the definition: $T=R_E/v_T$.
 
We adopt the model of an isothermal spherical halo in which the
normalized MACHO number distribution as a function of $v_T$ is
\begin{equation}
f(v_T) dv_T=\frac{2}{v_H^2} v_T e^{-v_T^2/v^2_H} dv_T~ , \label{eqno:4}
\end{equation}
with $v_H \approx 210$ km/s the velocity dispersion implied by the
rotation curve of our galaxy.
The MACHO number density distribution per unit mass $dn/d\mu$ is
given by
\begin{equation}
\frac{dn}{d\mu}=H(x)\frac{dn_0}{d\mu}=\frac{a^2+R^2_{GC}}
{a^2+R^2_{GC}+D^2x^2-2DR_{GC} x cos \alpha}~\frac{dn_0}{d\mu}, \label{eqno:5}
\end{equation}
with $dn_0/d\mu$ the local MACHO mass distribution.
We have assumed that $dn/d\mu$ factorizes in functions of $\mu$ and $x$
\cite{kn:Derujula}.
We take $a = 5.6$ kpc as the galactic ``core'' radius
(our final results do not depend much on the poorly known value of $a$),
$R_{GC} = 8.5$ kpc
our distance from the centre of the galaxy and $\alpha = 82^0$
the angle between the line of sight and the direction of the galactic
centre.
For an experiment monitoring $N_{\star}$ stars during a total observation
time $t_{obs}$ the number of expected microlensing events is
given by \cite{kn:Derujula}
\begin{equation}
N_{ev}=\int dN_{ev} =N_{\star} t_{obs}
2Dr_E \int v_T f(v_T) (\mu x(1-x))^{1/2} H(x) \frac{dn_0}{d\mu}
d\mu du_{min} dv_T dx   \label{eqno:6}
\end{equation}
where the integration variable $u_{min}$ is related to $A_{max}$:
$A_{max}=A[u = u_{min}]$.
For a more complete discussion 
in particular on the integration range see \cite{kn:Derujula}.

From eq.(\ref{eqno:6}) with some variable transformation
(see \cite{kn:Jetzer2})
we can define, up to a normalization constant,
the probability $P$ that a microlensing event of
duration $ T$ and  maximum amplification $A_{max}$
be produced by a MACHO of mass $\mu$, that we see first of all
is independent of $A_{max}$ \cite{kn:Jetzer2}
\begin{equation}
P(\mu,T) \propto \frac{\mu^2}{ T^4} \int_0^1 dx (x(1-x))^2 H(x)
exp\left( -\frac{r_E^2 \mu x(1-x)}{v^2_H  T^2} \right) ~. \label{eqno:8}
\end{equation}
We also see that $P(\mu, T)=P(\mu/ T^2)$. The measured values for
$ T$ are listed in Table 1, where 
$\mu_{MP}$ is
the most probable value.
We find that the maximum corresponds to 
$\mu r_E^2/v^2_H T^2=13.0$ \cite{kn:Jetzer2,kn:Jetzer1}. 
The 50\% confidence interval
embraces for the mass $\mu$ approximately
the range $1/3\mu_{MP}$ up to $3 \mu_{MP}$.
Similarly one can compute $P(\mu, T)$ also for the bulge events
(see \cite{kn:Jetzer1}).

\vskip 0.5cm 
Table 1: Values of $\mu_{MP}$ (in $M_{\odot}$)
for eight microlensing events detected in the LMC ($A_{i}$
= American-Australian
collaboration events ($i$ = 1,..,6);
$F_1$ and $F_2$
French collaboration events).
For the LMC: $v_H = 210~{\rm km}~{\rm s}^{-1}$ and
$r_E = 3.17 \times 10^9~{\rm km}$.

\begin{center}
\begin{tabular}{|c|c|c|c|c|c|c|c|c|}\hline
  & $A_{1}$ & $A_{2}$ & $A_3$ & $A_4$ & $A_5$ & $A_6$ & $F_1$ & $F_2$  \\
\hline
$ T$ (days) & 17.3 & 23 & 31 & 41 & 43.5 & 57.5 & 27 & 30 \\
\hline
$\tau (\equiv \frac{v_H}{r_E} T)$ & 0.099 &
0.132 & 0.177 & 0.235 & 0.249 & 0.329 & 0.155 & 0.172 \\
\hline
$\mu_{MP}$ & 0.13 & 0.23 & 0.41 & 0.72 & 0.81 & 1.41 & 0.31 & 0.38 \\
\hline
\end{tabular}
\end{center} 

\section{Mass moment method}

A more systematic way to extract information on the masses is to use the
method of mass moments as presented in De R\'ujula et al. \cite{kn:Derujula}. 
The mass moments $<\mu^m>$ are defined as
\begin{equation}
<\mu^m>=\int d\mu~ \epsilon_n(\mu)~ 
\frac{dn_0}{d\mu}\mu^m~. \label{eqno:10}
\end{equation}
$<\mu^m>$ is related to $<\tau^n>=\sum_{events} \tau^n$,
with $\tau \equiv (v_H/r_E) T$, as constructed
from the observations and which can also be computed as follows
\begin{equation}
<\tau^n>=\int dN_{ev}~ \epsilon_n(\mu)~
\tau^n=V u_{TH} \Gamma(2-m) \widehat H(m) <\mu^m>~,
\label{eqno:11}
\end{equation}
with $m \equiv (n+1)/2$ and
\begin{equation}
V \equiv 2 N_{\star} t_{obs}~ D~ r_E~ v_H=2.4 \times 10^3~ pc^3~ 
\frac{N_{\star} ~t_{obs}}{10^6~ {\rm star-years} }~, \label{eqno:12}
\end{equation}
\begin{equation}
\Gamma(2-m) \equiv \int_0^{\infty} \left(\frac{v_T}{v_H}\right)^{1-n}
f(v_T) dv_T~,
\label{eqno:13}
\end{equation}
\begin{equation}
\widehat H(m) \equiv \int_0^1 (x(1-x))^m H(x) dx~.  \label{eqno:14}
\end{equation}
The efficiency $\epsilon_n(\mu)$ is determined as follows 
(see \cite{kn:Derujula})
\begin{equation}
\epsilon_n(\mu) \equiv \frac{\int d N^{\star}_{ev}(\bar\mu)~ 
\epsilon(T)~ \tau^n}
{\int d N^{\star}_{ev}(\bar\mu)~ \tau^n}~, \label{eqno:15}
\end{equation}
where $d N^{\star}_{ev}(\bar\mu)$ is defined as $d N_{ev}$ 
in eq.(\ref{eqno:6}) with
the MACHO mass distribution concentrated at a fixed mass
$\bar\mu$: $dn_0/d\mu=n_0~ \delta(\mu-\bar\mu)/\mu$. 
For a more detailed discussion on the efficiency see ref.\cite{kn:Masso}.

A mass moment $< \mu^m >$ is thus related to 
$< \tau^n >$ as given from the measured values 
of $T$ in a microlensing experiment by
\begin{equation}
< \mu^m > = \frac{< \tau^n >}{V u_{TH} \Gamma(2-m) \hat H(m)}~.
\label{eqno:16}
\end{equation}

The mean local density of MACHOs (number per cubic parsec)
is $<\mu^0>$. The average local mass density in MACHOs is
$<\mu^1>$ solar masses per cubic parsec. 
In the following we consider only 6 (see Table 1)
out of the 8 events observed by the MACHO group,
in fact the two events we neglect are 
a binary lensing event and an event which is rated as marginal.
The  mean mass, which we get from
the six events detected by the MACHO team, is 
\begin{equation}
\frac{<\mu^1>}{<\mu^0>}=0.27~M_{\odot}~.
\label{eqno:aa}
\end{equation}
(To obtain this result we used the values of $\tau$
as reported in Table 1, whereas $\Gamma(1)\widehat H(1)=0.0362$ and
$\Gamma(2)\widehat H(0)=0.280$ as
plotted in figure 6 of ref. \cite{kn:Derujula}).
If we include also the two EROS events we get a value
of 0.26 $M_{\odot}$ for the mean mass.
The resulting mass depends on the parameters
used to describe the standard halo model. In order to check this
dependence we varied the parameters within
their allowed range and found
that the average mass changes at most by $\pm$ 30\%, which shows
that the result is rather robust. 
Although the value for the average mass we find with the mass moment
method is marginally consistent with the result of the MACHO team,
it definitely favours a lower average MACHO mass.

One can
also consider other models with more general
luminous and dark matter distributions, e.g. ones with a flattened halo
or with anisotropy in velocity space \cite{kn:Ingrosso},
in which case the resulting
value for the average mass would decrease significantly.
If the above value will be confirmed, then MACHOs cannot be brown dwarfs
nor ordinary hydrogen burning stars, since for the latter
there are observational
limits from counts of faint red stars. Then stellar remnants 
such as white dwarfs
are the most likely explanation. A scenario with white dwarfs
as a major constituent of the
galactic halo dark matter has been explored recently \cite{kn:Fields}.
However, it has some problems, since it requires that
the initial mass function
must be sharply peaked around $2 - 6 M_{\odot}$.
Given these facts, we feel that the brown dwarf option can still provide a 
sensible explanation of the above-mentioned microlensing events.
Notice also, that brown dwarfs have been discovered quite recently
in the solar neighbourhood and in the Pleiades cluster. 

Another important quantity to be determined is the fraction $f$ of the local
dark mass density (the latter one given by $\rho_0$) detected
in the form of MACHOs, which is given by
$f \equiv {M_{\odot}}/{\rho_0} \sim 126~{\rm pc}^3$ $<\mu^1>$.
Using the values given by the MACHO collaboration
for their two years data \cite{kn:Pratt} (in particular
$u_{TH}=0.661$ corresponding to $A > 1.75$ and
an effective exposure $N_{\star} t_{obs}$
of $\sim 5 \times 10^6$ star-years for 
the observed range of the event duration $T$ between $\sim$ 20 - 50 days)
we find $f \sim 0.54$, which compares quite well
with the corresponding value ($f \sim 0.45$ based on the six events
we consider) calculated
by the MACHO group in a different way. The value for $f$ is obtained 
again by assuming a standard spherical halo model. 

\vskip 0.2cm
Table 2: Values of
$\mu_{MP}$ (in $M_{\odot}$) as
obtained by the corresponding $P(\mu, T)$ for 
eleven microlensing events detected by OGLE
in the galactic bulge \cite{kn:Masso}. 
($v_H = 30~{\rm km}~{\rm s}^{-1}$ and
$r_E = 1.25 \times 10^9~{\rm km}$.) ($T$ is in days as above.)

\begin{center}
\begin{tabular}{|c|c|c|c|c|c|c|c|c|c|c|c|}\hline
  & 1 & 2 & 3 & 4 & 5 & 6 & 7 & 8 & 9 & 10 & 11\\
\hline
$ T$ & 25.9 & 45 & 10.7 & 14 &12.4& 8.4& 49.5&18.7&61.6&12&20.9 \\
\hline
$\tau$ & 0.054 &
0.093 & 0.022 & 0.029 & 0.026&0.017& 0.103& 0.039& 0.128& 0.025& 0.043\\
\hline
$\mu_{MP}$ & 0.61 & 1.85 & 0.105 & 0.18 &0.14& 0.065& 2.24& 0.32&
3.48 & 0.13 & 0.40 \\
\hline
\end{tabular}
\end{center}
\vskip 0.2cm

Similarly, one can also get information from the events
detected so far towards the galactic bulge.
The mean MACHO mass, which one gets when considering
the first eleven events detected by OGLE in the galactic bulge (see Table 2),
is $\sim 0.29 M_{\odot}$ \cite{kn:Jetzer1}.
From the 40 events discovered 
during the first year of operation
by the MACHO team \cite{kn:MACHO} (we considered
only the events used by the MACHO team to infer the optical
depth without the double lens event)
we get an average value
of 0.16$M_{\odot}$.
The lower value inferred from the MACHO data is due to the fact
that the efficiency for the short duration events ($\sim$ some days)
is substantially higher for the MACHO experiment than for the
OGLE one. 
These values for the average mass
suggest that the lens are faint disk stars. 

Once several moments $< \mu^m >$ are known one can
get information on the mass distribution $dn_0/d\mu$. 
Since at present only few events toward the LMC are at disposal the 
different moments (especially the higher ones) can 
be determined only approximately.
Nevertheless, the results obtained so far
are already of interest and it is clear that in a few years,
due also to the new experiments under way (such as EROS II and OGLE II),
it will be possible to draw more firm conclusions.

A major problem which arises is to explain the formation of MACHOs, as well
as the nature of the remaining amount of dark matter in the galactic halo.
We feel it hard to conceive a formation mechanism which transforms with 100\%
efficiency hydrogen and helium gas into MACHOs. Therefore, we expect
that also cold clouds (mainly of $H_2$) should be present in the 
galactic halo. Recently, we have proposed
a scenario \cite{kn:Roc,kn:Roncadelli}
in which dark clusters of MACHOs and cold molecular 
coulds naturally form in the halo at galactocentric distances
larger than 10-20 kpc, where the relative abundance depends on the
distance. 
We have also considered several observational tests for our model.
In particular, halo molecular clouds would produce a $\gamma$-ray flux through
the interaction with cosmic-ray protons. The detection of this $\gamma$-ray
flux is below current detectability but might be observed with
a new generation of gamma-ray satellites.




\end{document}